\documentclass[fleqn,usenatbib]{mnras}

\usepackage{newtxtext,newtxmath}

\usepackage[T1]{fontenc}
\usepackage{ae,aecompl}

\usepackage{graphicx}	
\usepackage{amsmath}	
\usepackage{amssymb}	
\usepackage{wasysym}    
\usepackage{url}        

\title[A Supermassive Black Hole in UCD3]{A 3.5-million Solar Masses Black Hole in the Centre of the Ultracompact Dwarf Galaxy Fornax UCD3}

\author[Afanasiev et al.]{Anton V. Afanasiev,$^{1,2}$\thanks{E-mail: anton.afanasiev@voxastro.org (AA),
igor.chilingarian@cfa.harvard.edu (IC)}
Igor V. Chilingarian,$^{3,1}$ Steffen Mieske,$^{4}$ Karina T. Voggel,$^{5}$ \newauthor
Arianna Picotti,$^{6}$ Michael Hilker$^{7}$, Anil Seth,$^{5}$ Nadine Neumayer,$^{6}$ Matthias Frank,$^{7}$
\newauthor
Aaron~J. Romanowsky,$^{8,9}$ George Hau,$^{4}$ Holger Baumgardt,$^{10}$ Christopher Ahn,$^{5}$
\newauthor
Jay Strader,$^{11}$ Mark den Brok,$^{12,13}$ Richard McDermid,$^{14}$ Lee Spitler,$^{15}$ Jean Brodie$^{9}$
\newauthor 
and Jonelle L. Walsh$^{16}$.
\\
\\
$^{1}$Sternberg Astronomical Institute, M.V. Lomonosov Moscow State University, 13 Universitetsky prospect, Moscow, 119234, Russia\\
$^{2}$Department of Physics, M.V. Lomonosov Moscow State University, 1, Leninskie Gory, Moscow, 119234, Russia \\
$^{3}$Smithsonian Astrophysical Observatory, 60 Garden St. MS09, Cambridge, MA, 02138, USA\\
$^{4}$European Southern Observatory, Alonso de Cordova 3107, Vitacura, Santiago, Chile\\
$^{5}$Department of Physics and Astronomy, University of Utah, 115 South 1400 East, Salt Lake City, UT 84112, USA \\
$^{6}$Max Planck Institute for Astronomy, K\"onigstuhl 17, D-69117 Heidelberg, Germany\\
$^{7}$European Southern Observatory, Karl-Schwarzschild-Str. 2, 85748 Garching bei M\"unchen, Germany\\
$^{8}$Department of Physics and Astronomy, San Jos\'{e} State University, San Jose, CA 95192, USA\\
$^{9}$University of California Observatories, 1156 High Street, Santa Cruz, CA 95064, USA\\
$^{10}$School of Mathematics and Physics, The University of Queensland, St. Lucia, QLD 4072, Australia\\
$^{11}$Department of Physics and Astronomy, Michigan State University, East Lansing, MI 48824, USA\\
$^{12}$ETH Z\"urich, Wolfgang-Pauli-Strasse 27, CH-8093 Z\"urich, Switzerland\\
$^{13}$Leibniz-Institut f\"ur Astrophysik Potsdam, AIP, An der Sternwarte 16, D-14482 Potsdam, Germany\\
$^{14}$Department of Physics and Astronomy,  Macquarie University, NSW 2109, Australia\\
$^{15}$Australian Astronomical Observatory, P.O. Box 915, North Ryde, NSW 1670, Australia\\
$^{16}$George P. and Cynthia Woods Mitchell Institute for Fundamental Physics and Astronomy, Department of Physics and Astronomy, \\
Texas A\&M University, College Station, TX 77843, USA\\
}

\date{Accepted XXX. Received YYY; in original form 28.02.2018}

\pubyear{2018}

\begin{document}
\label{firstpage}
\pagerange{\pageref{firstpage}--\pageref{lastpage}}
\maketitle

\begin{abstract}
The origin of ultracompact dwarfs (UCDs), a class of compact stellar systems discovered two decades ago, still remains a matter of debate. Recent discoveries of central supermassive black holes in UCDs likely inherited from their massive progenitor galaxies provide support for the tidal stripping hypothesis. At the same time, on statistical grounds, some massive UCDs might be representatives of the high luminosity tail of the globular cluster luminosity function. Here we present a detection of a $3.3^{+1.4}_{-1.2}\times10^6\,M_{\odot}$ black hole ($1\sigma$ uncertainty) in the centre of the UCD3 galaxy in the Fornax cluster, that corresponds to 4~per~cent of its stellar mass. We performed isotropic Jeans dynamical modelling of UCD3 using internal kinematics derived from adaptive optics assisted observations with the SINFONI spectrograph and seeing limited data collected with the FLAMES spectrograph at the ESO VLT. We rule out the zero black hole mass at the $3\sigma$ confidence level when adopting a mass-to-light ratio inferred from stellar populations. This is the fourth supermassive black hole found in a UCD and the first one in the Fornax cluster. Similarly to other known UCDs that harbour black holes, UCD3 hosts metal rich stars enhanced in $\alpha$-elements that supports the tidal stripping of a massive progenitor as its likely formation scenario. We estimate that up to 80~per~cent of luminous UCDs in galaxy clusters host central black holes. This fraction should be lower for UCDs in groups, because their progenitors are more likely to be dwarf galaxies, which do not tend to host central black holes.
\end{abstract}




\section{Introduction}  
Ultracompact dwarfs (UCDs) are a class of compact stellar systems discovered about two decades ago during spectroscopic surveys of the Fornax cluster. The first such object that looked like an ultra-luminous star cluster mentioned in \citet{Minniti98} was claimed to ``represent the nucleus of a dissolved nucleated dwarf elliptical galaxy'' \citep{Hilker99} and later got the name UCD3 \citep{Drinkwater00}. The typical size of UCD is between $10$ and $100$ pc (e.g. ~\citealp{Drinkwater03,Hasegan05,Mieske06,Hilker07}) with masses varying from several millions to a few hundred million solar masses (e.g.~\citealp{Mieske13}). This puts them among the densest stellar systems in the Universe, along with nuclear clusters \citep{Misgeld11,Walcher05,Norris15}.
In the mass--size plane, UCDs occupy a sequence connecting the largest globular clusters (GCs) with compact elliptical galaxies (cE) \citep{Norris14}. 
Given their intermediate nature between 'normal' star clusters and dwarf galaxies, the proposed formation scenarios of UCDs articulate around two main channels: UCDs may be the most massive star clusters (e.g. ~\citealp{Fellhauer02,Fellhauer05,Mieske02,Mieske12,Mieske17}), or be the tidally stripped nuclear remnants of massive dwarf or low-mass giant galaxies (\citealp{Phillips01,Bekki03,Pfeffer13,Strader13}).

An interesting observational result of UCD studies in the last decade has been that their dynamical mass-to-light ratios appear to be systematically elevated with respect to canonical stellar population expectations (\citealp{Hasegan05,Dabring09,Dabring10,Dabring12,Baumgardt08,Mieske08K,Taylor10,Frank11,Strader13}). This finding has prompted suggestions of a stellar initial mass function (IMF) variation in UCDs (top-heavy: ~\citealp{Murray09,Dabring09,Dabring10}; bottom-heavy: ~\citealp{Mieske08K,Villaume17}).

Apart from IMF variations, it has also been proposed that central massive black holes, as relics of tidally stripped UCD progenitor galaxies, would cause such elevated global M/L ratios (e.g. ~\citealp{Mieske13}). The expected supermassive black hole (SMBH) masses required for the level of observed M/L increase are around 10--15~per~cent of the UCD masses. For three massive (> $10^7 M_{\odot}$) Virgo Cluster UCDs, such a dynamical detection of an SMBH signature have indeed been found through dynamical modeling of adaptive optics (AO) kinematics derived from AO assisted integral field unit (IFU) spectroscopy \citep{Seth14,Ahn17}. The relative SMBH masses derived from those measurements are in the expected mass range, making up $\sim12--18$~per~cent of the UCDs dynamical mass, and the resulting M/L values are in agreement with those derived from stellar population models. This may imply that the most massive UCDs with masses above $\sim 5 \times 10^7 M_{\odot}$ are indeed dominated by tidally stripped nuclei. UCDs with masses below $\sim 10^7 M_{\odot}$ are expected, on statistical grounds, to be the high-mass tail of the regular star cluster population \citep{Mieske12,Pfeffer14,Pfeffer16}. Although, based on the masses of known nuclear star clusters, and the $\sim$1 million $M_{\odot}$ nucleus of the tidally disrupting Sgr dwarf galaxy (e.g. \citealp{Siegel07}), it is also clear that at least some stripped nuclei should exist in this mass range.

While SMBHs have been confirmed in three Virgo Cluster UCDs, this is not yet the case for UCDs in the Fornax cluster, where UCDs were originally discovered. Tracing the SMBH occupation fraction in UCDs as a function of environment is important to gauge the relative importance of the various proposed UCD formation channels, and also to constrain the SMBH volume density in the local Universe \citep{Seth14}. \citet{Frank11} investigated the velocity dispersion profile of the most massive Fornax UCD, UCD3, based on good seeing (0.5~arcsec FWHM) ground-based IFU data. With this data set they derived a 2$\sigma$ upper limit of a possible SMBH mass in this UCD that corresponds to 20~per~cent of the UCD's mass. This upper limit still allows for an SMBH of the 10--15~per~cent mass range typically expected if the elevated average M/L ratios of UCDs are due to a central black hole. Given that seeing limited data even under good conditions do not allow such a measurement to be performed at Fornax cluster distance, one requires adaptive optics assisted spectroscopy. 

Fornax UCD3 has a compact and relatively bright ($V=18$~mag) nuclear component that is unresolved in ground-based imaging. This makes it feasible for laser guide star (LGS) AO assisted spectroscopic observations because it is bright enough to be used as a source for tip-tilt corrections. UCD3 has roughly solar metallicity and is slightly $\alpha$-enhanced with $[\alpha/\rm Fe]\approx 0.2$ \citep{Chilingarian11}. It is located at projected distance of $11~\rm kpc$ from the neighbouring giant elliptical galaxy NGC~1404, in the central part of the Fornax galaxy cluster. However, the radial velocity difference between UCD3 and NGC 1404 is $430 \rm~km~s^{-1}$, while other nearby giant elliptical galaxy NGC1399 located $50 \rm~kpc$ of projected distance away from UCD3 has its radial velocity different only by $90 \rm~km~s^{-1}$. We assume a distance of $20.9~\rm Mpc$ to this galaxy \citep{Blakeslee09}. UCD3 has an apparent magnitude magnitude $m_V=18.06$, total magnitude $M_V=-13.33$ and average color $\mu_{F606W}-\mu_{F814W}=0.65$ \citep{Evstigneeva08}. 

In this paper we present LGS AO observations of Fornax UCD3 collected with the Spectrograph for INtegral Field Observations in the Near Infrared (SINFONI, \citealp{Eisenhauer03,Bonnet04}) and the dynamical modelling of UCD3 based on these data. Our goal is to obtain a SMBH mass estimate for UCD 3 with a sensitivity well below the SMBH mass range of 10--15~per~cent expected from indirect arguments and also found for three Virgo cluster UCDs. The paper is structured as follows: in Section~\ref{spectra} we report on the SINFONI AO data obtained for Fornax UCD3, in Section~\ref{HST} we describe the HST imaging data used for the analysis, in Section~\ref{JAM} we present the dynamical modelling and its results, and in Section~\ref{disc} we discuss our findings.

\begin{figure}
\includegraphics[width=\hsize]{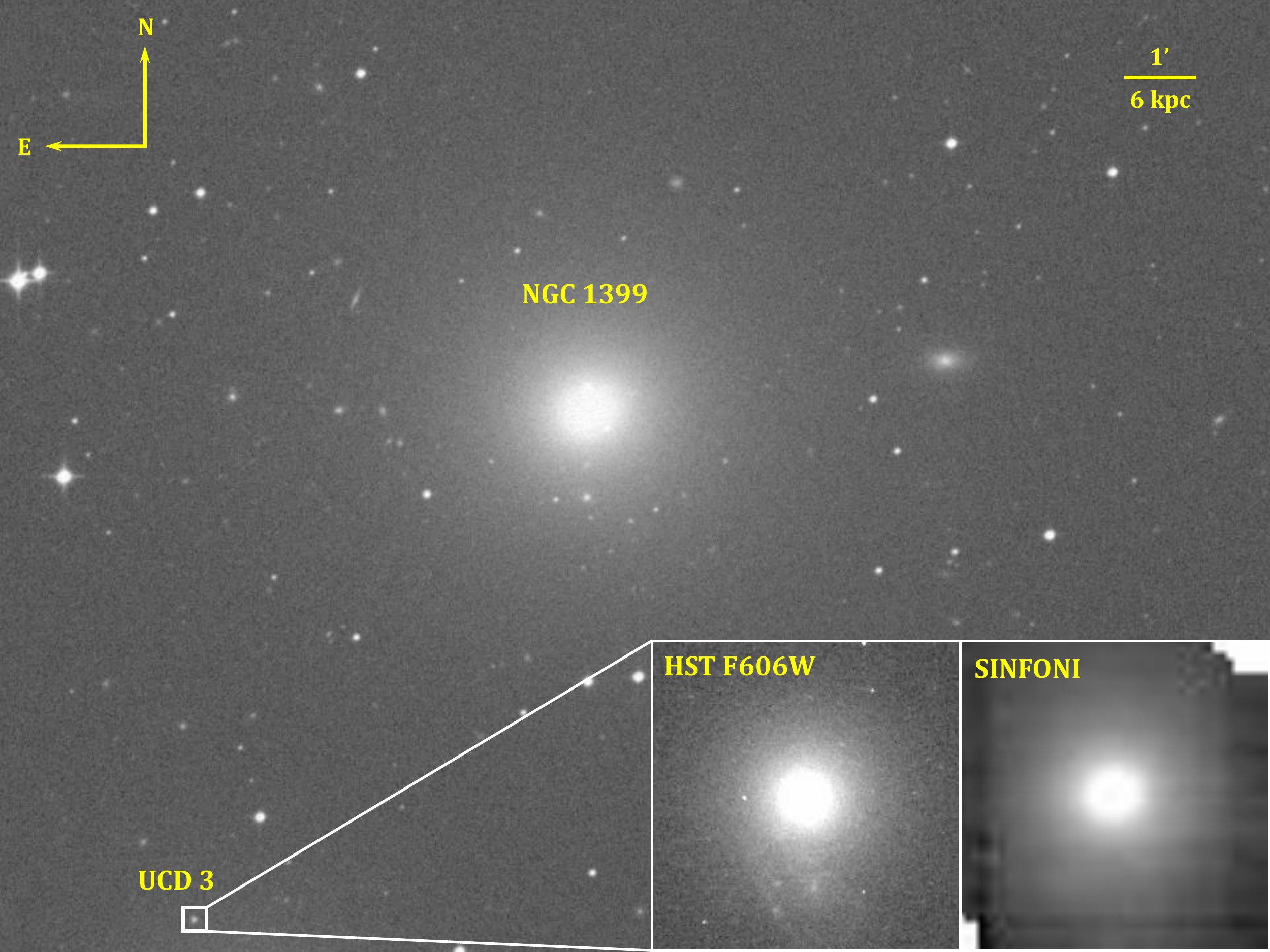}
\caption{\label{galaxy_overview} An archival $Spitzer~Space~Telescope$ image of the central region of the Fornax cluster demonstrating the location of UCD3 near the central galaxy NGC~1399. The insets show: (a) an $HST$ ARC/HRC image of UCD3 in the F606W band, (b) a reconstructed image from the SINFONI datacube in the $K$ band. Both insets are displayed in the same spatial scale and measure 1~arcsec $\approx$ 101~pc on a side.}
\end{figure}

\section{Spectroscopic Observations, Data Reduction and Analysis}\label{spectra} 

\subsection{SINFONI observations.}
UCD3 was observed as a part of the program 095.B-0451(A) (P.I.: S. Mieske) with the SINFONI IFU spectrograph operated at the Cassegrain focus of the European Southern Observatory Very Large Telescope UT4 at cerro Paranal, Chile.  
SINFONI is a cryogenic integral field spectrograph providing adaptive optics assisted low- and intermediate-resolution spectroscopy in the $J$, $H$, and $K$ bands.  
For our program we observed in the K-band (2.0-2.4 um; R~3500) with a spatial scale of 0.1~arcsec per spaxel and a resulting field of view of 3x3~arcsec and used a laser guide star to operate the VLT adaptive optics system.  
The compact core of UCD3 served as a tip-tilt star.

Observations were split into 7 observing blocks (OBs) and performed in service mode during the nights of 2015/Aug/25 (1 OB), 2015/Sep/15 (3 OBs), and 2015/Sep/16 (3 OBs).  
Each observing block included three 10-min long on-source science exposures in two dithering positions with two 10-min long offset sky positions (object, sky, object, sky, object).  
The natural seeing quality was about 0.8~arcsec during the first and third night, and 0.6~arcsec during the second night.  
Hence, the total on-source integration time was 3h~30min.  
Thanks to the adaptive optics system, the spatial resolution of the combined dataset was around 0.18~arcsec (the inner PSF peak containing 58~per~cent of the light), see below.

\subsection{Data reduction and post-processing.}

\begin{figure}
\includegraphics[width=0.60\hsize]{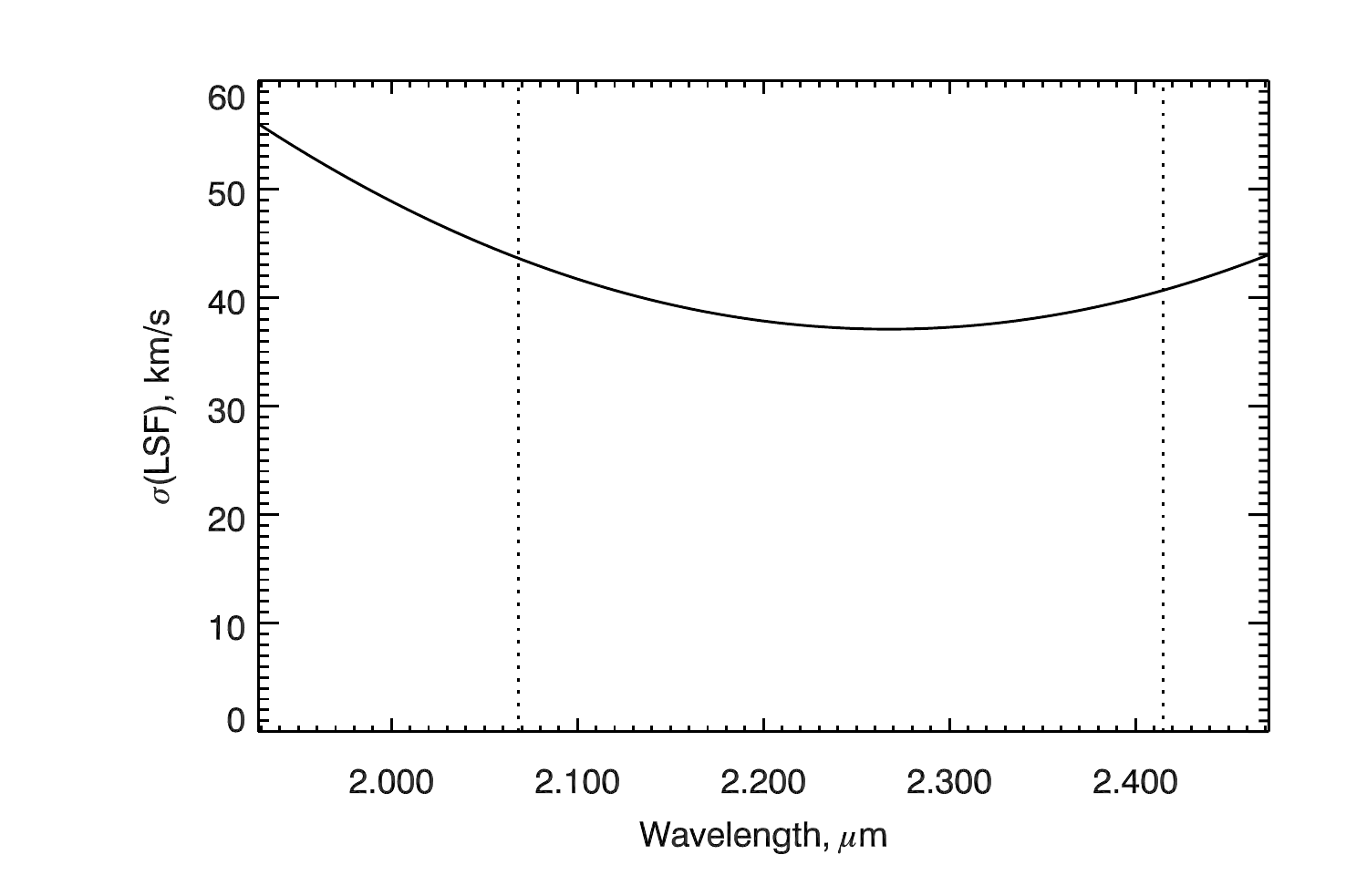}
\includegraphics[width=0.39\hsize]{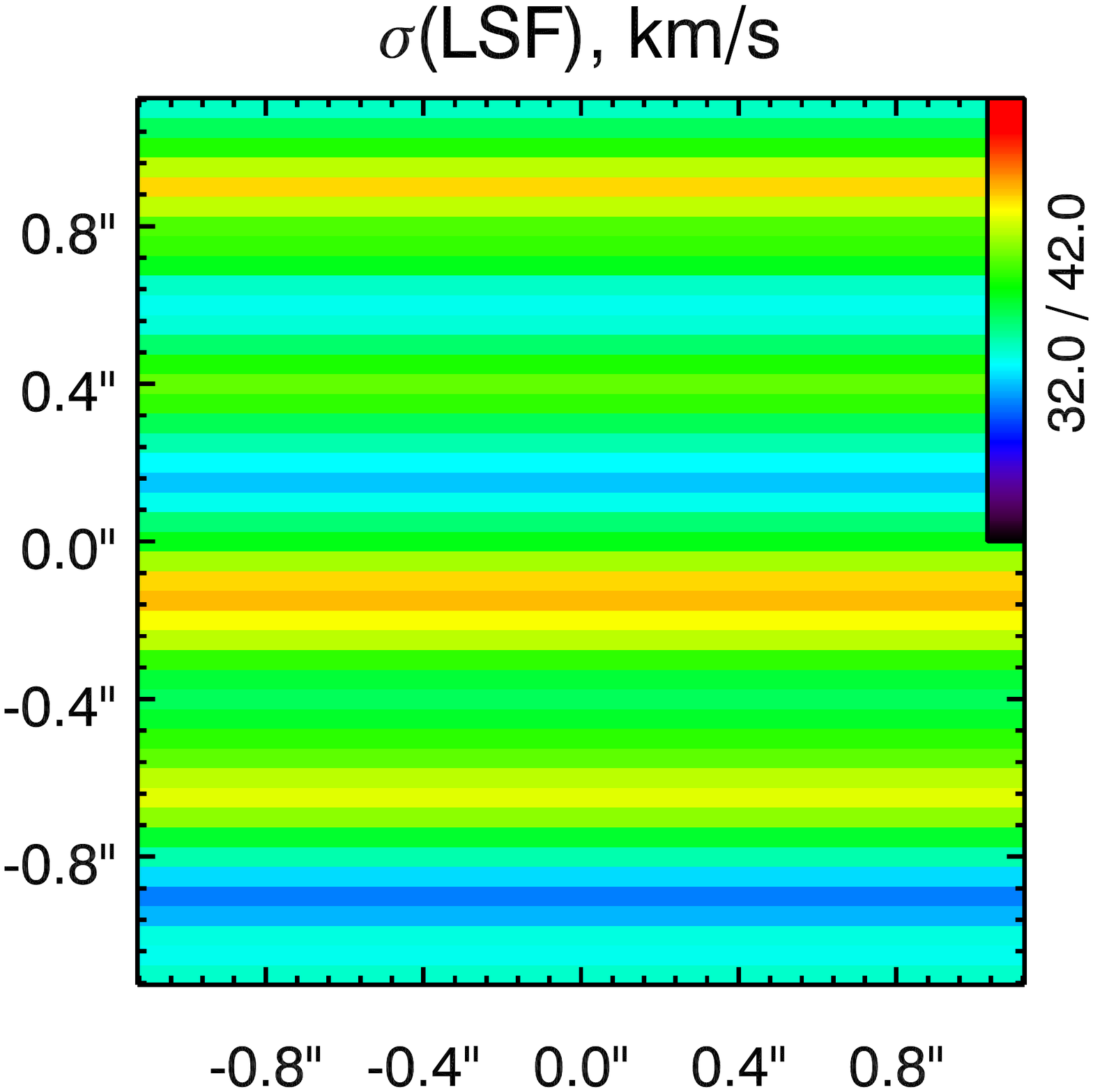}
\caption{\label{lsf_figure} The average spectral line spread function
variations of SINFONI along the wavelength (average trend, left panel) and across the field of view in the 2$\times$2~arcsec region centered on UCD3 (right panel) derived from the co-added non-sky-subtracted datacube.}
\end{figure}

We used the standard ESO SINFONI data reduction pipeline {\sc sinfo/2.7.0} for the preliminary data reduction. The pipeline created cosmic ray rejected spatially over-sampled datacubes for each observing block with the spatial scale of 0.05~arcsec~pix$^{-1}$. However, the inspection of reduced datasets revealed significant residuals around atmospheric hydroxil (OH) and water vapor emission lines. Also, the standard telluric correction procedure using a telluric standard star did not yield satisfactory results. Therefore, we had to perform heavy post-processing of original pipeline products.

(i) In each reduced observing block we identified a vertical stripe in the outer region of the field of view, which did not include any significant contribution from UCD3. We estimated the residual background using the outlier resistant mean along the $x$ direction in that region at every wavelength and then subtracted it from the entire field of view at that wavelength. The reason for one-dimensional averaging is the specific pattern of the spectral line spread function variation across the field of view (see below). 

(ii) We determined the UCD position in a datacube resulting from each
observing block by fitting a two-dimensional Gaussian into the synthetic
image computed by collapsing a datacube along the wavelength. Then we resampled observing blocks 2--7 spatially using bi-linear interpolation in order to match the UCD3 position in the first OB. The same operation was performed on non-sky-subtracted datacubes.  

(iii) We co-added spatially resampled datacubes and obtained co-added
datasets; the flux uncertainties were computed as square root of
non-sky-subtracted flux: our observations were all background dominated (photon noise level $>15 e^-$), therefore the read-out noise part ($\sim3.5 e^-$) could be neglected.

(iv) Then we fitted several single OH lines in non-sky-subtracted datacube by Gaussians and determined variations of the SINFONI spectral line spread function (LSF) across the field of view and along the wavelength.  The spatial variations have a very specific pattern with well defined horizontal stipes with the instrumental Gaussian $\sigma$ ranging from 33 ($R\approx3800$) to 42~km~s$^{-1}$ ($R\approx3000$). $\sigma$ has its minimum at 2.25~$\mu$m increasing by about 35~per~cent at shorter (2.00~$\mu$m) and 13~per~cent at longer (2.45~$\mu$m) wavelengths. The variations of the LSF Gaussian width are shown in Fig~\ref{lsf_figure}.

(v) We used the nuclear region of UCD3 integrated in a 0.2~arcsec aperture in order to estimate the telluric correction in a manner similar to what is implemented in the MMT and Magellan Near-Infrared Spectrograph data reduction pipeline \citep{Chilingarian+15} with a few modifications: (a) we fitted an observed UCD3 spectrum by a linear combination of template stellar spectra (M and K giants) broadened with a Gaussian line-of-sight velocity distribution (LOSVD); the templates were multiplied by the atmospheric transmission model obtained by the interpolation of a model grid computed for different water vapor and airmass values with the airmass value fixed at the average airmass during the observation, all convolved with the LSF computed at the previous step; (b) we used the best-fitting atmospheric transmission model convolved with the LSF at every position in the field of view to correct the entire dataset. This algorithm will be described in detail in a forthcoming paper presenting data reduction pipelines for Magellan optical and near-infrared Echelle spectrographs MagE and FIRE.

\subsection{Full spectrum fitting.}

\begin{figure}
\includegraphics[width=\hsize]{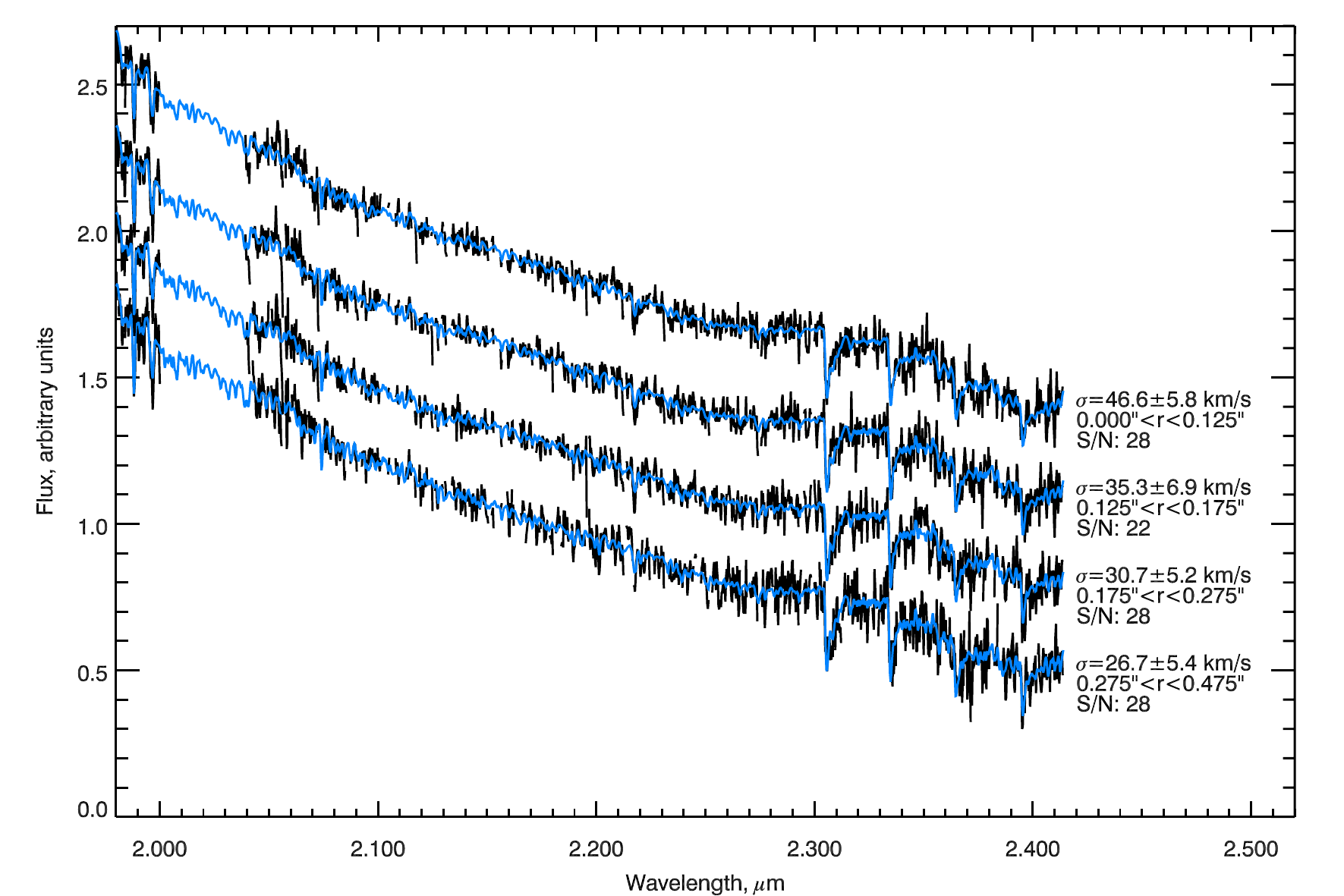}
\caption{SINFONI spectra of UCD3 in radial bins (black) and their
best-fitting stellar templates (blue). The best-fitting values of stellar
velocity dispersions, inner and outer radii of bins and the resulting
signal-to-noise values at 2.25$\mu$m are shown near each spectrum. The
spectra are offset in the vertical direction for display
purposes.\label{fig_spectra}}
\end{figure}

Having obtained a fully reduced telluric corrected datacube and maps of the LSF variations, we extracted the galaxy kinematics using the penalized pixel fitting code \citep{Cappellari04} by fitting galaxy spectra against stellar templates. We binned the datacube into 4 annular bins having the mean radii between 0.038~arcsec and 0.42~arcsec which yielded the signal-to-noise ratios per pixel at 2.25~$\mu$m between 22 and 30 in every bin (see Fig.~\ref{fig_spectra}). Then we used a grid of 15 synthetic templates from the latest generation of stellar atmospheres computed with the PHOENIX code \citep{Husser13} for the following atmosphere parameters: $T_{\mbox{eff}}=3200 K, \log g = 1.0$, $T_{\mbox{eff}}=3600 K, \log g = 1.0$, $T_{\mbox{eff}}=4000 K, \log g = 1.5$, $T_{\mbox{eff}}=4400 K, \log g = 1.5, 2.0$ for the three values of metallicities $[Fe/H]=-0.5, 0.0, +0.5$~dex and solar $\alpha$/Fe abundance ratios. These synthetic atmospheres are representative of M and K giants at a large metallicity range.  We used the wavelength range from 1.98 to 2.41~$\mu$m excluding the area between 2.00 and 2.04~$\mu$m heavily affected by telluric absorptions which could not be fully corrected. Finally, we obtained velocity dispersion measurements in 5 radial bins in an overall range between 26 and 53 km/s (see Fig.~\ref{fig_spectra}), with uncertainties
of about 5--7~km~s$^{-1}$. 

In Fig.~\ref{fig_spectra} we present the 4 spectra and their best-fitting
templates. The most prominent features are the 4 $CO$ bands visible at
wavelengths between 2.3 and 2.4~$\mu$m. The central bin has a significantly higher velocity dispersion value (53$\pm$7~km~s$^{-1}$) than the outermost ones (26--27~km~s$^{-1}$). The radial velocities in all 5 bins are consistent within statistical errors. Given the very low amplitude of rotation detected in UCD3 by \citet{Frank11} ($v\approx 3$~km~s$^{-1}$) and relatively low signal-to-noise of our data, we decided not to split our annular bins in the azimuthal direction.  We did run a test, which did not detect any statistically significant rotation in 4 azimuthal bins with the outer radius of 0.15~arcsec. In order to check whether our measurements are subject to wavelength calibration errors across the entire wavelength coverage, we also extracted velocity dispersions from a short spectral region (2.28$<\lambda<2.40 \mu$m), which contains CO bands. We obtained the values full consistent with those extracted from the full spectral range, however, having somewhat higher uncertainties because the number of pixels was smaller and the average signal-to-ratio was also lower.

UCD3 has a significantly $\alpha$-enhanced stellar population \citep{2009MNRAS.394.1801F}. 
Therefore, we also tried to include $\alpha$-enhanced stellar atmosphere
models into the grid, which we used in the fitting procedure. However, they always ended up with zero weights in a linear combination when used together with the Solar scaled models. This may indicate some imperfections in the stellar atmosphere modelling, but this discussion stays beyond the scope of our study. We stress, however, that if we use $\alpha$-enhanced templates only, the resulting radial velocity values do not change significantly, but the resulting $\chi^2$ increase thus suggesting slightly worse overall fitting quality.

As a consistency test with the published data, we also extracted a velocity dispersion value in the central circular region ($r<0.3$~arcsec) which should roughly correspond to the central value obtained by \citet{Frank11} from the seeing limited optical observations (27$\pm$0.5~km~s$^{-1}$. Our value, $\sigma_{\mathrm{0.6"}}=33.0\pm4.7$~km~s$^{-1}$ is $1.4\sigma$ higher, which we attribute to the image quality difference between our AO assisted observations and the seeing limited data. 

\section{Analysis of Archival Hubble Space Telescope Images} \label{HST}
\subsection{Surface photometry and Multiple Gaussian Expansion}
In order to create precise luminosity and mass models of UCD3, we need imaging data of the highest spatial resolution available. In this work we used archival Hubble Space Telescope data from the HST snapshot program 10137 (PI M.~Drinkwater). The data were taken using the High Resolution Channel (HRC) on the Advanced Camera for Survey (ACS), which provides a pixel scale of $0.025$~arcsec~$\rm{pixel}^{-1}$. We used data in two filters, $F606W$ and $F814W$ with the corresponding exposure times of $870$~s and $1050$~s. The PSFs for these images were modeled using the TinyTim\footnote{\url{http://tinytim.stsci.edu/}} software \citep{Krist93}.  

We subtracted the sky background and performed a two-dimensional light profile decomposition for UCD3 using the {\sc galfit~3} software package \citep{Peng10}. We tested two ways of fitting the background level, using a tilted plane and an extended S\'{e}rsic profile centered on the neighbouring giant galaxy NGC~1404. The UCD3 light profile decomposition results were consistent within the statistical errors in both cases. 

We fitted galaxy images in both filters with a PSF-convolved model, which included two S\'ersic components. All parameters (centre positions, S\'ersic indices, effective radii, luminosities, axial ratios $b/a$, and position angles) were allowed to vary, although it did not lead to sufficient centre mismatch or unreasonably low $b/a$ relation. Both components were found to be close to circular, with $b/a > 0.9$ and centre position difference less than one pixel. We found the best-fitting parameters in $F606W$ band to be $R_e=0.177$~arcsec, $n=1.43$, $m_{tot}=20.17$ for the inner component and $R_e=2.269$~arcsec, $n=1.55$, $m_{tot}=17.66$ for the outer component. In the $F814W$ band, the best-fitting model parameters turned out to be $R_e=0.308$~arcsec, $n=2.08$, $m_{tot}=19.00$ and $R_e=2.315$~arcsec, $n=1.04$, $m_{tot}=17.19$. We notice that the southern outskirts of UCD3 are projected onto a background spiral galaxy which could in principle affect the quality of the photometric decomposition, but in practice there are no noticeable inconsistencies of the fitting results between the two photometric bands.

As a next step, in order to prepare the input data for the dynamical modelling, we need to compute an Multi-Gaussian Expansion (MGE) of the galaxy light profile \citep{Cappellari02}. The original routine by M.~Cappellari allows one to derive an MGE directly from a galaxy image. While being able to yield precise results when applied to galaxies extended over hundreds of pixels, in the case of UCD3, the object is too small to have enough data points along the radial extent of the surface brightness distribution, and moreover, it is affected by the presence of a background galaxy on one side (see Fig.~\ref{galaxy_overview}). Therefore, we used a different technique that  derives MGE Gaussians from our best-fitting two component S\'{e}rsic profiles (see \citealp{Seth14}). 

\subsection{UCD3 colour gradient -- is it real?}
For the purpose of dynamical modelling, we need to convert the MGE profile into stellar densities and, therefore we need to know the stellar $M/L$ ratio. It is typically assumed that mass follows light, so the MGE gaussians can simply be multiplied by a constant. However, we cannot adopt this approach if the galaxy possesses significant colour variations along the radius. 

\citet{Evstigneeva08} reported that the UCD3 colour profile $F606W-F814W$ has a ``step'' of $\sim0.2$~mag~$\rm{arcsec}^{-2}$ at the radius $\sim0.15$~arcsec, i.e. the inner part is bluer than the outer envelope. Also, this color gradient can be explicitly seen if one calculates the synthetic colour using two-component S\'{e}rsic models not convolved with the PSF. However, the unconstrained photometric fitting results imply that the inner component has a colour $\mu_{F606W} - \mu_{F814W}=1.17$~mag and the outer component has a colour $\mu_{F606W} - \mu_{F814W}=0.47$~mag. That leads to an apparent contradiction, because the inner component looks redder than the outer one, despite the bluer central colour in the overall colour profile.

On the other hand, the effective radius and S\'{e}rsic index values of the inner component are different in $F606W$ and $F814W$ bands ($R_{F606W}=0.177$~arcsec, $R_{F184W}=0.308$~arcsec; $n_{F606W}=1.43$, $n_{F814W}=1.55$). This difference might originate from the degeneracy between the effective radius and the S\'{e}rsic index. In order to verify how this might affect our analysis, we refitted the $F814W$ band data by fixing the inner S\'{e}rsic index to the $F606W$ band value ($n=1.43$), and then reconstructed the colour profile from the two-component S\'{e}rsic models in the two bands.
\begin{figure}
\center{\includegraphics[width=1\linewidth]{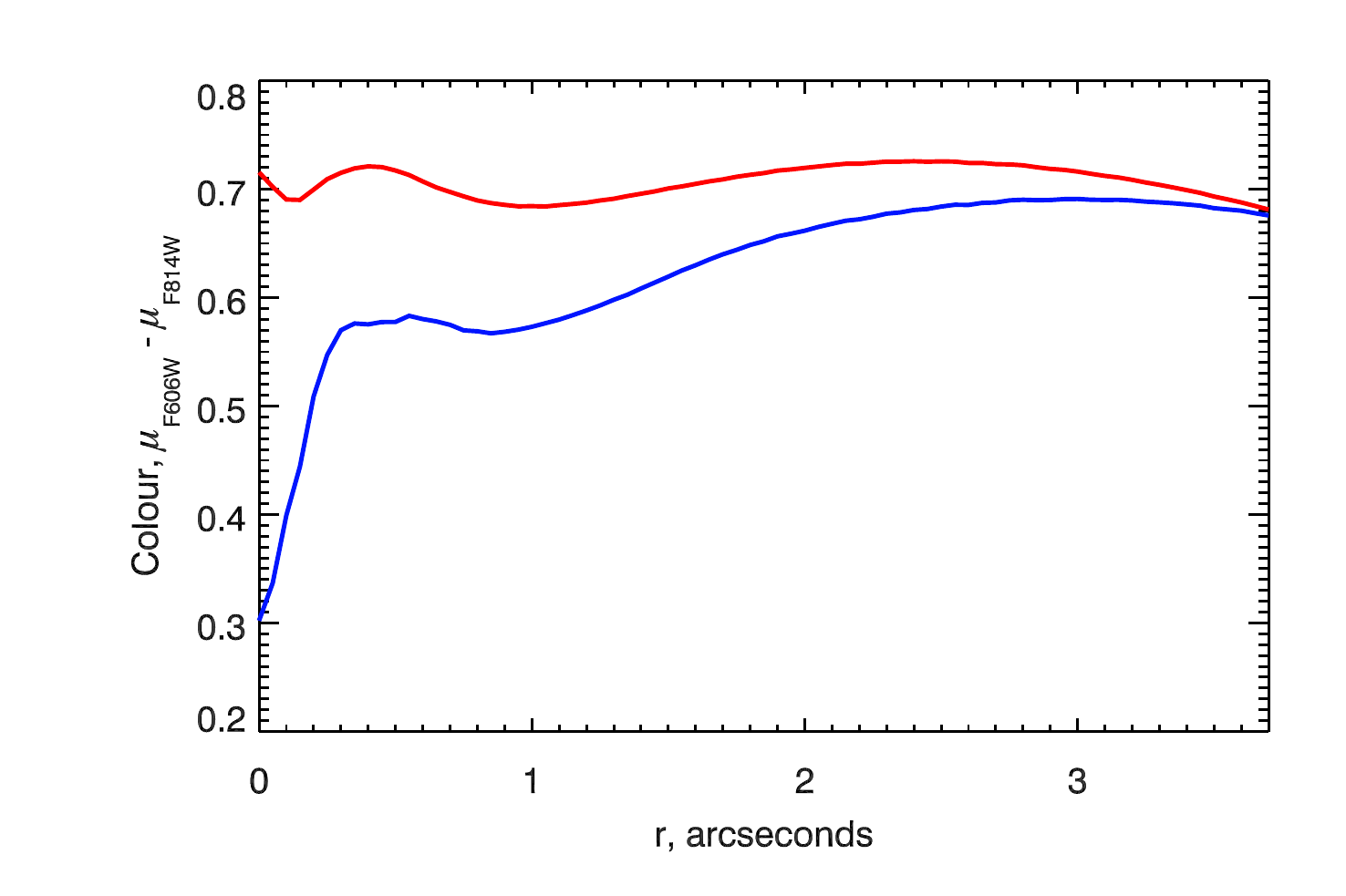}}
\caption{Colour profile of UCD3. The blue line is colour profile of synthetic two-component S\'{e}rsic model with free parameters. The red line shows synthetic profile if the S\'{e}rsic index in $F814W$ band is fixed to the best-fitting value in $F606W$.}
\label{fig:colour}
\end{figure}
Then in the $F814W$ band, the best-fitting model parameters become $R_e=0.192$~arcsec, $n_{\mathrm{fix}}=1.43$, $m_{tot}=19.57$~mag for the inner component and $R_e=2.296$~arcsec, $n=1.48$, $m_{tot}=17.03$~mag for the outer component respectively. This model yields the $\chi^2/\rm{DOF}$ value similar to the unconstrained fitting ($0.13115$ vs $0.13128$). And at the same time, it eliminates the strong colour gradient towards the centre, making the reconstructed colour profile almost flat with overall variations less than $\pm 0.03$~mag and the component average colour difference of only $0.02$~mag (see Fig.~\ref{fig:colour}). 

Such a flat colour profile supports the ``mass follows light'' assumption, so we can now use the constant $M/L$ ratio for the entire mass model. This also suggests that in $K$-band the relative intensities of the inner and outer S\'{e}rsic components should not be very different compared to the $F606W$ and $F814W$ bands. Finally, we adopted the structural parameters obtained from the $F606W$ band photometric data for the dynamical modelling procedure because they have higher spatial resolution compared to the $F814W$ band and, therefore, we expect the derived values for the S\'ersic components are more reliable. 
We adopt the stellar $(M/L)_{V}=3.7 (M/L)_{\odot}$ from \citet{Chilingarian11} for our analysis, but we also compute a grid of models leaving $M/L$ as a free parameter. According to predictions by {\sc pegase.2} \citep{FR97}, the expected colour $V_{\mathrm{Johnson}}-F606W$ for a 13~Gyr old stellar population is $\sim 0.04$~mag. For the subsequent analysis we need to derive the $M/L$ ratio in the $F606W$ band. The $(M/L)_{F606W}$ estimate is obtained from $(M/L)_{V}$ by combining the $V-F606W$ colour of UCD3 and the difference in the absolute magnitudes of the Sun in the same bands (0.08~mag). This, we adopt the value $(M/L)_{F606W}=3.35$. In this fashion, we obtained the photometric MGE profile for UCD3 in the $F606W$ band and transformed it into stellar densities. The final multi-Gaussian expansion of the Fornax-UCD3 light profile converted into stellar densities is presented in Table~\ref{tab:MGE}. 

\begin{table}
\begin{tabular}{|c|c|c|c|c|}
\hline
\hline
$M_{\odot} pc^{-2}$ & $L_{\odot} pc^{-2}$ & $\sigma$ & $\sigma$ &  $q$  \\
 & & (arcsec) & (pc) & \\
\hline
       26903    &    8030.8  & 0.0008415   &  0.08526   &   0.99 \\
       37113    &    11079   &  0.002815   &   0.2852   &   0.99 \\
       41134    &    12279   &  0.007744   &   0.7847   &   0.99 \\
       35382    &    10562   &   0.01825   &    1.849   &   0.99 \\
       22662    &    6764.7  &   0.03794   &    3.844   &   0.99 \\
       10303    &    3075.5  &   0.07128   &    7.222   &   0.99 \\
       3190.7   &    952.44  &    0.1231   &    12.47   &   0.99 \\
       655.37   &    195.63  &    0.1989   &    20.15   &   0.99 \\
       82.891   &    24.744  &    0.3085   &    31.26   &   0.99 \\
       4.4744   &    1.3356  &    0.4857   &    49.21   &   0.99 \\
       542.13   &    161.83  &   0.02157   &    2.185   &   0.95 \\
       1029.5   &    307.33  &   0.07762   &    7.865   &   0.95 \\
       1422.7   &    424.69  &    0.2054   &    20.81   &   0.95 \\
       1333.6   &    398.08  &    0.4402   &    44.61   &   0.95 \\
       770.95   &    230.13  &    0.8033   &    81.40   &   0.95 \\
       257.88   &    76.980  &     1.302   &    132.0   &   0.95 \\
       45.059   &    13.450  &     1.959   &    198.6   &   0.95 \\
       2.7506   &   0.82106  &     2.892   &    293.1   &   0.95 \\

\hline
\end{tabular}
\caption{Mass and  luminosity model of UCD3 composed of 18 Gaussians based on the \textit{F606W}-band HST ACS image. Col. (1):  galaxy mass density. Col. (2): galaxy \textit{F606W}-band surface brightness.  Col. (3) and Col. (4): size along the major axis. Col. (5): axis ratio.\label{tab:MGE}}
\end{table}

\section{Anisotropic Jeans Model of UCD3} \label{JAM}
\subsection{Jeans anisotropic modelling}
Here we analyse the kinematic and photometric data for UCD3 using the Jeans Anisotropic Modelling (JAM, \citealp{Cappellari08}). The code implementing the JAM method and the method itself are discussed in detail in the same paper, here we provide only a brief outline of the approach. The JAM method relies on the two basic assumptions: (1) the velocity ellipsoid of a galaxy is aligned to the axes of a cylindrical coordinate system, i.e. it is axisymmetric (2) the anisotropy profile is constant with radius. Here we specify anisotropy as a vertical anisotropy $\beta_z=1-\left( \frac{\sigma_z}{\sigma_R}\right)^2$, where $\sigma_z$ is the velocity dispersion along the rotation axis and $\sigma_R$ is the radial dispersion in the plane of rotation of the galaxy. The procedure of deriving radial velocity and velocity dispersion models consists of several steps. First, it creates a 3D mass (or light) model by deprojecting 2D MGE Gaussian components obtained from the photometric data. In the second step it generates the gravitational potential using that mass distribution. This potential also contains a central Gaussian representing a supermassive black hole. Then the MGE formalism is applied to the solution of axisymmetric anisotropic Jeans equations. Finally, the 3D model decomposed by MGE components is integrated along the line-of-sight and convolved with the spatial PSF of the kinematic observations that yields 2D model maps of stellar radial velocity and velocity dispersion \citep{Cappellari12_JAM}.

We computed the spatial PSF for our SINFONI data using the following steps. (i) We collapsed the co-added SINFONI datacube along the wavelength excluding the regions around atmospheric OH airglow and H$_2$O lines. (ii) We resampled the background subtracted \textit{HST} ACS HRC image in the $F606W$ band to match the scale of the reconstructed image. (iii) We used {\sc galfit} and fitted a two-component PSF model composed of two circular Gaussians to a reconstructed SINFONI image using the \textit{HST} image as a ``PSF'' . The best-fitting kinematic PSF model derived in this fashion includes the inner component with a FWHM of $0.173$~arcsec containing $64$~per~cent of light and the outer component with a  FWHM of $0.503$~arcsec containing $36$~per~cent of light. 

Now we can compute JAM-based predictions of radial velocity and velocity dispersions in UCD3 for different values of the central black hole mass, anisotropy, inclination, and mass-to-light ratio, and compare it to the observed kinematic data by means of the standard $\chi^2$ statistics. 

\citet{Frank11} demonstrated that Fornax-UCD3 has a very slow global rotation with a maximal observed velocity $\sim3$~km~s$^{-1}$. This value is too low to be confidently detected using our SINFONI data given the uncertainty of our velocity measurements of 7--9~km~s$^{-1}$ and the outermost bin radius of 0.4~arcsec. Because of the low signal-to-noise ratio (see Section~2) we analysed only the average kinematical profiles rather than 2-dimensional maps. In order to extend the velocity dispersion profile to 1.5~arcsec, we added 3 outermost data points from \citet{Frank11}, because at $r>0.5$~arcsec we can neglect the influence of the atmospheric seeing. Assuming the maximal rotation velocity of UCD3 is only $\sim3$~km~s$^{-1}$, we will use velocity root-mean-square ($RMS=\sqrt{v^2+\sigma^2}$) models to fit our velocity dispersion profile, as their difference never exceeds $1$~per~cent. 

Masses of central supermassive black holes are often estimated using dynamical modelling methods, which allow for fully general distribution functions (e.g. \citealp{Schwarzschild79}). This can be important, because central dispersion peaks can be explained either by anisotropy variations or by the presence of a black hole. Highly eccentric radial stellar orbits have an average radius far from the centre of a galaxy and high pericentric velocities of stars, thus raising the observed central velocity dispersion in a similar way to what happens when a supermassive black hole is present, but without a significant increase of the central mass density. However, in the case of UCD3 we will adopt an isotropic solution because of nearly face-on orientation of UCD3, abundance of isotropic systems among compact axisymmetric objects and some other reasons discussed in detail in Section~\ref{sec:anisotropy}. 

\subsection{Dynamical modelling results}
\begin{figure}
\center{\includegraphics[width=1\linewidth]{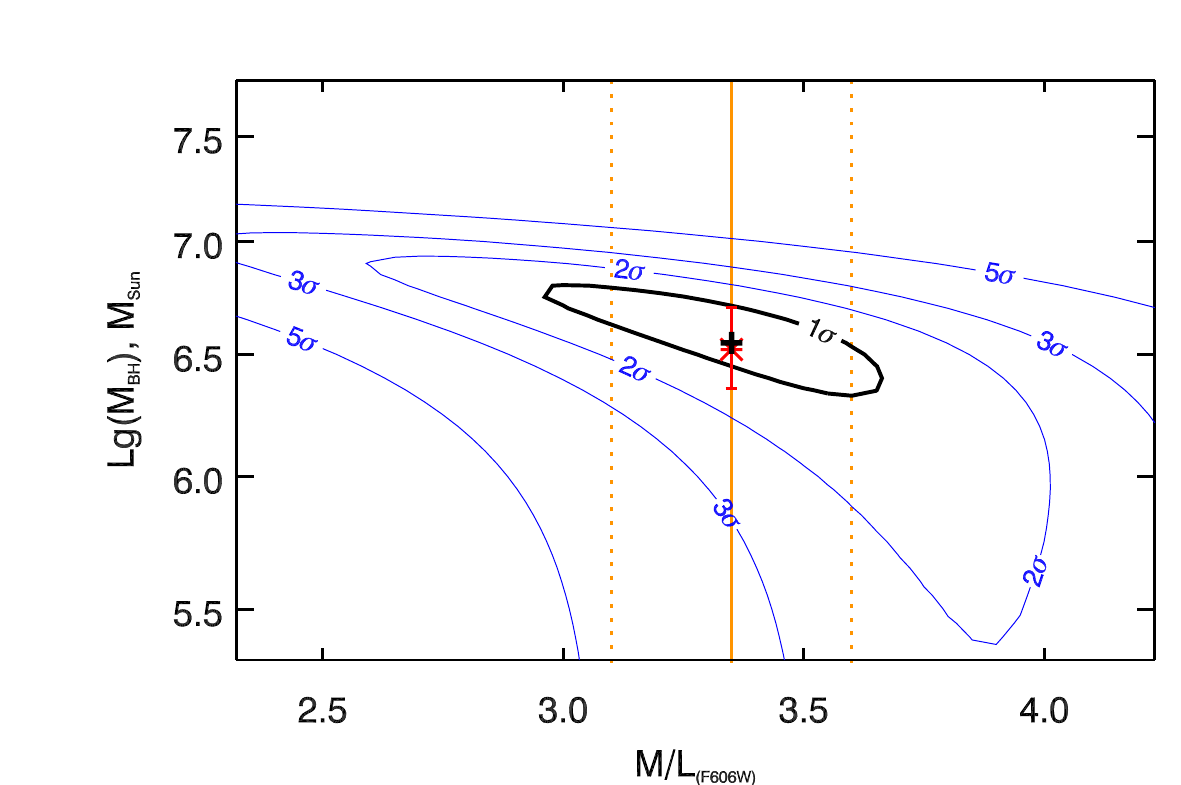}}
\caption{The contours of black hole mass and $M/L$ determination uncertainties. $1-\sigma$ contour is solid black, $2-$, $3-$ and $5-\sigma$ error contours are shown by blue lines. The minimum of $\chi^2$ statistic is marked by the black cross and is located at $M_{BH}=3.5\times 10^{6}$ and $M/L=3.35$. The solid orange vertical line shows $M/L=3.35$ determined in \citet{Chilingarian11}, while the dashed orange lines show the uncertainty of this estimate. The red star shows the best-fitting black hole mass obtained with fixed best-fitting $M/L=3.35$ using cumulative likelihood function, and the red error bars demonstrate the $1\sigma$ error of this estimate. The significance of non-zero mass black hole for $M/L=3.35$ is confidently higher than $3\sigma$, while the overall black hole significance is $2\sigma$. The inclination for this map was set to $19.8 \rm~deg$}
\label{fig:chimap}
\end{figure} 

At first, we attempted to use spherically symmetric Jeans models, because the axial ratio of MGE Gaussians $q=0.95$ brings them close to circular. However the fitting results of spherical models against observational data were found unsatisfactory. If we try to yield a reasonable $M/L$ lying within our $(M/L)_{F606W}$ estimate uncertainties, the spherical models will be inconsistent with both FLAMES and SINFONI data (see details in Figure~\ref{fig:profile_apdx}). These data could in principle be fitted by a spherical model, but it will result in unreasonably low $M/L\lesssim2$, high $M_{BH}\sim8\times 10^{6} M_{\odot}$, and a $\chi^2$ statistic significantly higher than in the axisymmetric case. Hence, we decided to use Jeans axisymmetric models for the subsequent analysis and add the galaxy rotation plane inclination as a free parameter.

\begin{figure}
\center{\includegraphics[width=1\linewidth]{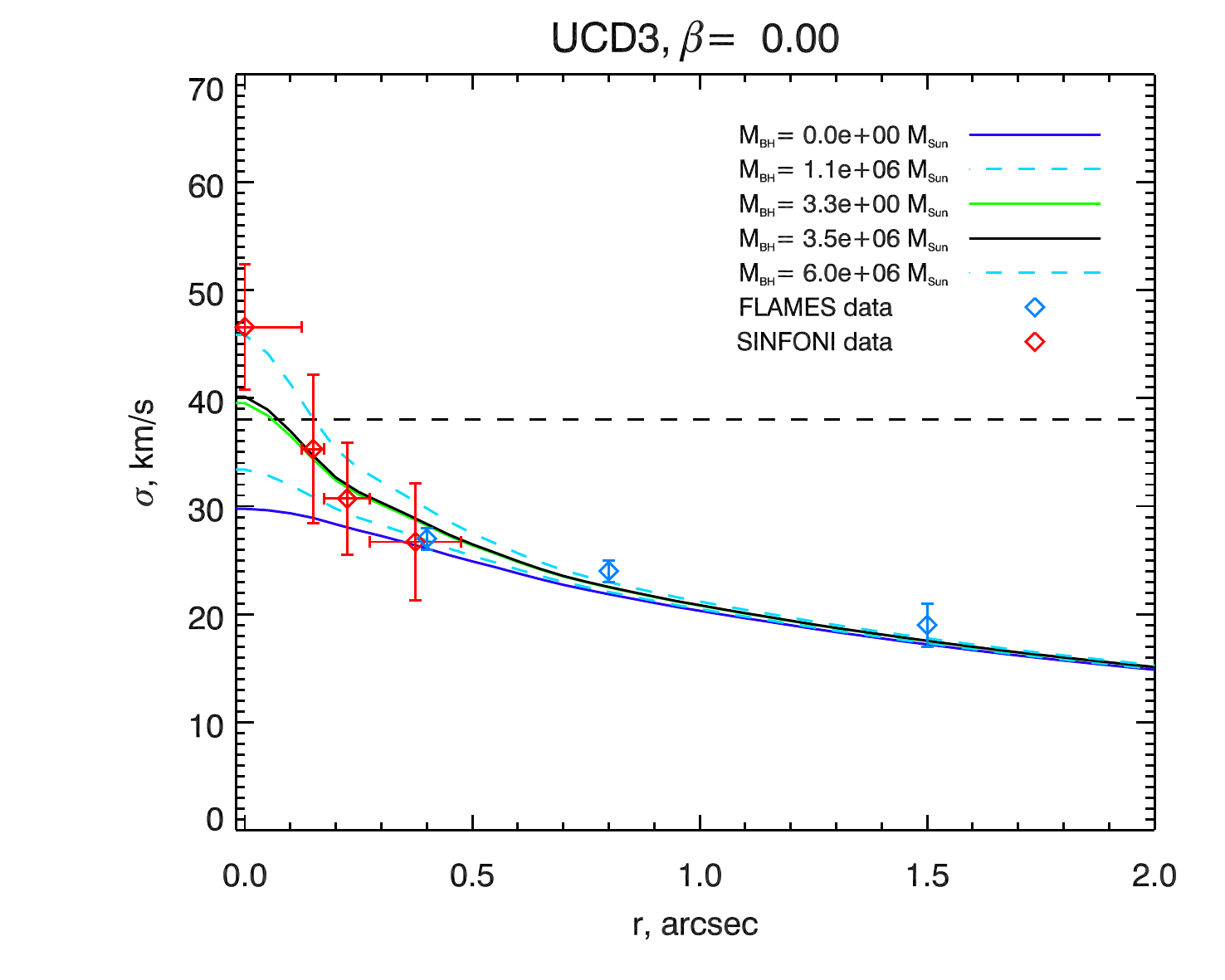}}
\caption{The results of the isotropic dynamical modelling of UCD3. The green solid line indicates the best-fitting black hole mass, with dashed cyan lines representing the 2-$\sigma$ uncertainties of the mass determination. The blue line shows the best-fitting isotropic model without a black hole. The blue dots indicate dispersion data obtained with the FLAMES spectrograph by \citet{Frank11}. The red dots indicate our data from \textit{SINFONI}. The horizontal error bars of SINFONI data demonstrate the radial binning of our spectroscopic data. The dashed horizontal line is the average $\sigma_{\rm LSF}$ of SINFONI. The dispersion rise in the center clearly exceeds $\sigma_{LSF}$, while the curve corresponding to $3.5\times 10^6 M_{\odot}$ is the closest to the majority of the data points. The $(M/L)_{V}$ for all models is fixed and set to $3.35$.}
\label{fig:profile}
\end{figure}

Now we present the results of fitting our velocity dispersion data to the isotropic ($\beta_z=0$) axisymmetric dynamical models (see models in Figure~\ref{fig:profile}). In order to estimate the black hole mass, $M/L$ ratio and inclination $i$ we computed the $\chi^2$ statistics in each point of the 2-dimensional $M_{BH} - M/L$ parameter grid for several values of inclination and plotted $1-$, $2-$, $3-$ and $5-\sigma$ confidence levels on that grid. The inclination values were chosen from the range $19\div24$~deg. In Figure~\ref{fig:chimap} we demonstrate a $\chi^2$ map for a best-fitting inclination $i=19.8$~deg. The explanation for such range and choice of best-fitting value is provided in Section~\ref{sec:anisotropy}. The $\chi^2$ minimum corresponds to the black hole mass of $3.5^{+3}_{-2}\times 10^6 M_{\odot}$ and $M/L=3.35\pm 0.4$. The overall probability of zero mass black is $2.3$~per~cent which implies a $2\sigma$ detection. However such a scenario would require a $(M/L)_{F606W}=3.9$ which is inconsistent with \citealp{Chilingarian11} estimate from stellar population analysis.

\begin{figure}
\center{\includegraphics[width=1\linewidth]{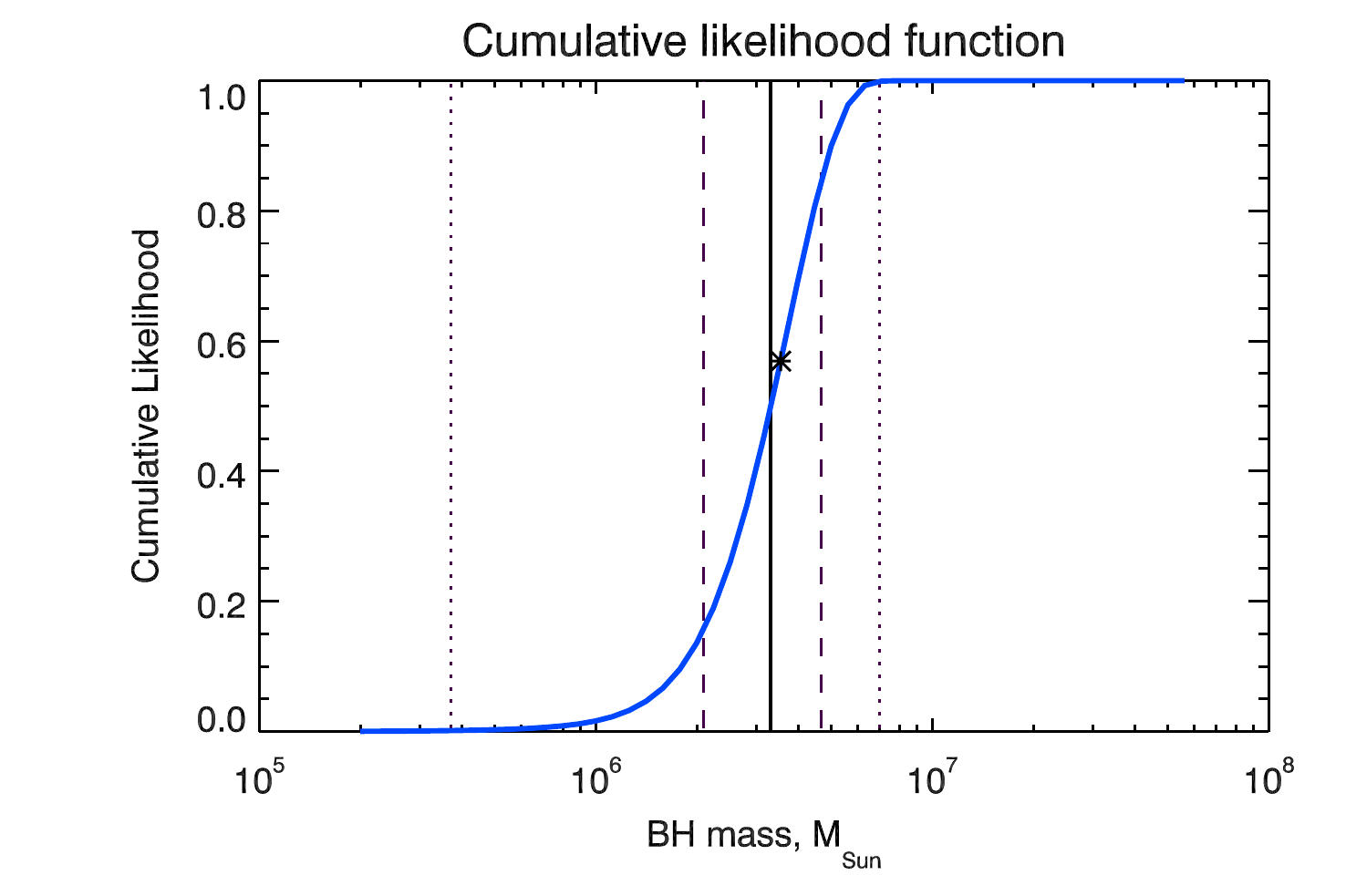}}
\caption{The cumulative likelihood function (blue) shows a clear rise beginning from $M_{BH}=2\times10^6 M_{\odot}$. The $1-$ and $3-\sigma$ errors are shown by vertical dashed and dotted lines respectively. The most probable black hole mass is marked with solid vertical line and is estimated as $3.32\times 10^6 M_{\odot}$. The black star marks $M_{BH}=3.5\times 10^6 M_{\odot}$, which yields minimal $\chi^2$. This cumulative likelihood function is calculated using $M/L=3.35$, and inclination $i=19.8 \rm~deg$. }
\label{fig:clf}
\end{figure}

Hereafter, we calculated the cumulative likelihood function (CLF) for the black hole mass using constant best-fitting $M/L=3.35$ (Fig.~\ref{fig:clf}), which resulted in most likely black hole mass value of $3.32^{+1.4}_{-1.2}\times 10^6 M_{\odot}$. Here we can rule out the zero mass black hole case with a confidence of $3\sigma$. We also present the results of the isotropic dynamical modelling of UCD3 for a best-fitting M/L ratio $M/L = 3.35~M_{\odot}/L_{\odot}$, $i=19.8$~deg and several central black hole mass values in Figure~\ref{fig:profile} to demonstrate the quality of fitting the kinematic data. However all obtained parameters are subject to some systematic errors and effects of degeneracy, which are discussed in relation to the dynamical estimate in Section~\ref{sec:anisotropy}.

\subsection{Influence of anisotropy, inclination and systematic errors} \label{sec:anisotropy}

The almost round shape of the isophotes in UCD3 as well as a low projected rotational velocity can be considered as evidence that the inclination of the rotational plane is low, i.e. that UCD3 is oriented nearly face-on if it rotates. The inclination lower limit ($19$~deg) is determined by the lowest intrinsic axis ratio $\rm{q}=0.95$ of our MGE Gaussians (see Table~\ref{tab:MGE}), and the upper limit is motivated by consistency with \citet{Frank11} velocity measurements. The modelling of the rotation velocity distribution showed that only inclination values under 24.5 degrees could reproduce the low observed rotation velocity. Such an orientation leads to the weak dependence of our model on the value of the vertical anisotropy. The central velocity dispersions in the models with $\beta_z=-100$ and $\beta_z=0.8$ differ by only $4$~km~s$^{-1}$, and more substantial differences can only be found with very extreme values of anisotropy. Thus, we fixed this parameter in our analysis and set the value of anisotropy to $\beta_z=0$.

\begin{figure}
\center{\includegraphics[width=1\linewidth]{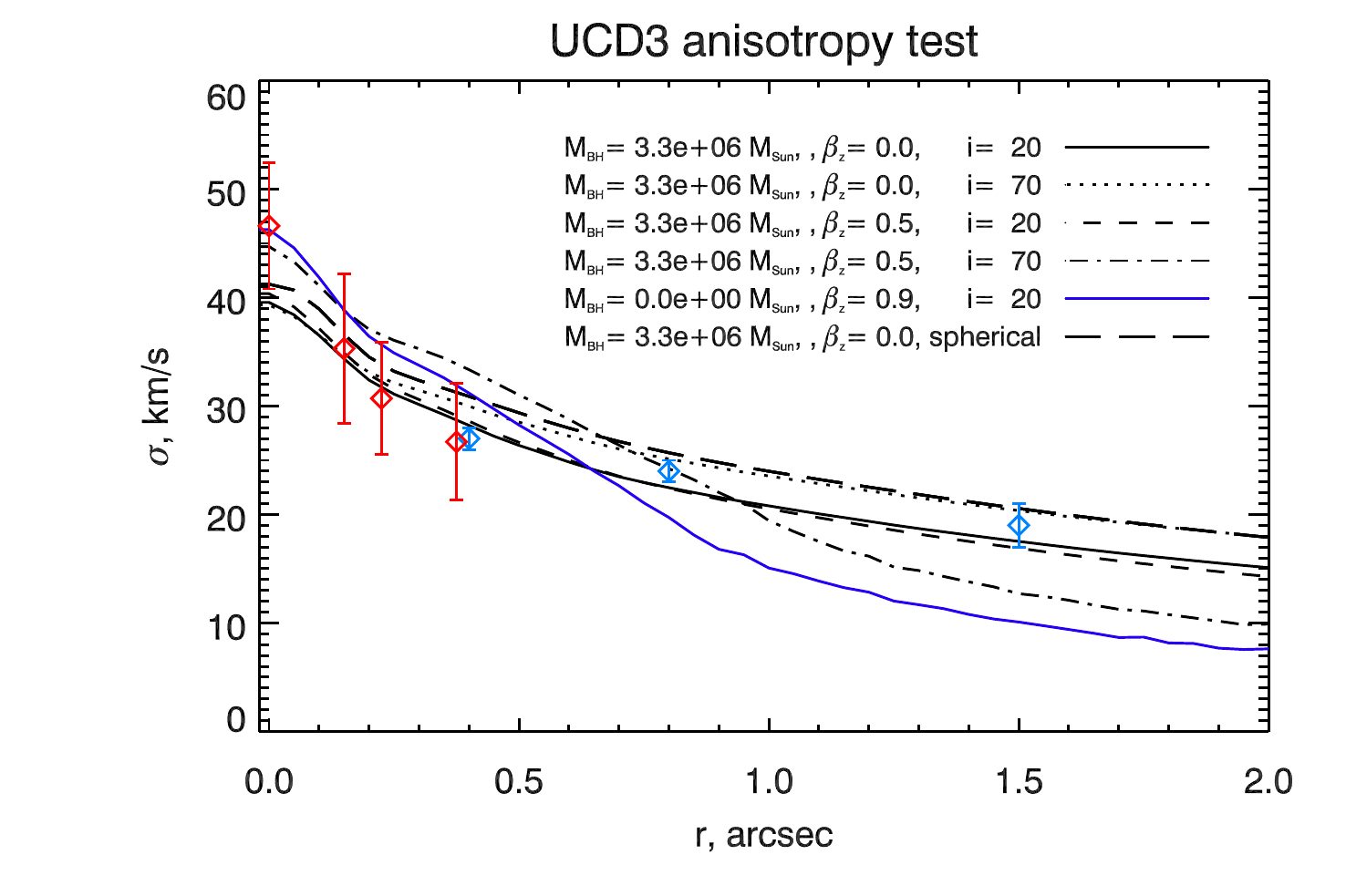}}
\caption{The results of the dynamic modelling with different values of anisotropy, inclination, and black hole mass. Black solid line shows the isotropic model with $M_{BH}=3.3\times10^6 M_{\odot}$, $(M/L)_V=3.35$ and $i=20^{\circ}$. The dotted line indicates isotropic model with increased inclination $i=70^{\circ}$. The dashed line represents model with $i=20^{\circ}$ but with anisotropy of $\beta_z=0.5$. The dashed-dotted line shows model with both inclination and anisotropy increased ($i=70^{\circ}$ and $\beta_z=0.5$). The blue line shows model with zero black hole mass and extreme anisotropy $\beta_z=0.9$. The long-dashed line represents the spherical model with the same parameters ($M_{BH}$, $M/L$, $\beta_z$) as the black solid line. The blue and the red data points are same as in Fig.~\ref{fig:profile}. All models assume $M/L=3.35$.}
\label{fig:profile_apdx}
\end{figure}

We investigate the possible influence of anisotropy on our resuts by analysing how the models change if we assume different values of $\beta_z$. The first test was done by assuming a black hole mass to be zero and exploring which value of $\beta_z$ would result in a rise of the velocity dispersion in the center comparable to the observed value of $\sim46$~km~s$^{-1}$. Assuming $M/L=3.35$ and $i=20^{\circ}$, this value turned out to be $\beta_z=0.9$ which seems completely unphysical as this would mean all stellar orbits in UCD3 are extremely eccentric. The value of anisotropy needed to reproduce the central dispersion of the best-fitting isotropic model, which peaks at about $42$~km~s$^{-1}$, is slightly lower and equals $\beta_z=0.85$. Still, such anisotropy is too high to be considered as a probable source of the central velocity dispersion peak, because such high anisotropy values have not been found in any observed systems including globular clusters \citep{Watkins15}, and galaxies \citep{Cappellari13}. Moreover, the dispersion profile with extremely high anisotropy falls off very fast in the outer regions, reaching $10$~km~s$^{-1}$ at $r=1.5$~arcsec (see Fig.~\ref{fig:profile_apdx}), which is completely inconsistent with the FLAMES observational data \citep{Frank11}.

The assumption of isotropic stellar orbits seems rather reasonable for compact stellar systems. In the case of M60-UCD1 where the quality of kinematic data was much higher than what we have for UCD3 because of the higher signal-to-noise ratio, the results of isotropic Jeans modelling were fully consistent with the results of the more sophisticated Schwarzschild modelling \citep{Seth14}. Moreover, nuclear structures in nearby galaxies are shown to be nearly isotropic \citep{Verolme02,Cappellari09,Schoedel09, Feldmeier14, FK17} and this property should be inherited by UCDs during the process of tidal stripping because it does not affect the mass distribution in the central region of a progenitor galaxy, similarly to what happens in more massive compact elliptical galaxies \citep{Chilingarian+09,Chilingarian15}. Finally, as it turns out from our tests, the vertical anisotropy $\beta_z$ is not a crucial parameter for the velocity dispersion profile shape in UCD3. Therefore, we decided to omit the precise evaluation of anisotropy and use isotropic models.

Variations in inclination affect the velocity dispersion profile in a more complex way. A 10-degree difference in inclination does not significantly change the inner region of the velocity dispersion profile, but the outer region is subject to a significant shift. The difference between the models with inclinations $i=19$~deg and $i=24$~deg is equivalent to the $M/L$ increase from $3.5$ to $5.5$. That results in the degeneracy between inclination and $M/L$ ratio in our $\chi^2$ minimization. The change in inclination from $19$~deg to $24$~deg results in shift of our best-fitting $M/L$ value from $M/L=4.2$ to $M/L=2.8$ with best-fitting $M_{BH}$ ranging from $2.5$ to $4$ million $M_{\odot}$. Hence, there is no possible way to distinguish $M/L$ and inclination effects on our models, unless we impose one of the two. We decided to adopt the inclination of $19.8~deg$ because it results in consistency of our $M/L$ estimates with $M/L$ evaluation from stellar population analysis. 

In Figure~\ref{fig:profile_apdx} we demonstrate the influence of the vertical anisotropy $\beta_z$ and inclination $i$ on the shape of model profiles. One can see that our model is sensitive to the variations in inclination, which seem to be degenerate with $M/L$. Both models with higher $i$ are inconsistent with outer data points \citep{Frank11}. The anisotropy tends to influence the model at minimal inclination very little, although if the inclination is increased, the difference between isotropic and anisotropic models grows promptly. Low anisotropy ($\lesssim 0.5$) is generally consistent with our modelling. Assuming such a value of anisotropy slightly increases the best-fitting $M/L$, but it cannot decrease black hole mass by more than $\approx 5-10$~per~cent.

Systematic errors also can influence the accuracy of the obtained result. The presence of a background spiral galaxy on the \textit{HST} image could influence both the quality of multi-Gaussian decomposition of UCD3 photometric profile and the estimate of the outer axial ratio. Thus the dynamical models and the minimal inclination are uncertain to some degree. A significantly stronger effect that impacts our black hole mass estimates is uncertainty in kinematic PSF determination. \citet{Ahn17} demonstrated that it is the dominant systematic error in that study, because the spatial resolution needed for confident black hole detection requires FWHM of PSF inner component to be narrower than atmospheric seeing. A $0.05$~arcsec increase in FWHM of PSF inner component changes the dynamical models significantly, making them fit the data much worse. The apparent inability of best-fitting JAM models to perfectly fit the highest ($46$~km~s$^{-1}$) data point is most likely connected with kinematic PSF uncertainties. The other source of systematic errors can be a procedure of annular binning of the kinematic data. Using our SINFONI observations we were able to obtain good constraints and reliable fitting results from a radial dispersion profile, but a 2D map was impossible to construct given the limited S/N ratio of the data. The variations of inner and outer radius of each bin led to a moderate change in the average velocity dispersion in that bin, and hence the change of the $\chi^2$ statistic minimal value and the best-fitting black hole mass. We countered this issue by comparing different binning options by the value of $\chi^2$ statistic at its minimum and chose the one which yields the minimal $\chi^2$ among all minimums. None the less, the good agreement between data and models does not necessarily mean that our binning represents exactly the original velocity dispersion profile.

\section{Discussion} \label{disc}

We detect the kinematic signature of a 3.5-million $M_{\odot}$ black hole in the centre of the ultracompact dwarf galaxy Fornax UCD3. This detection strongly supports the scenario that UCD3 was formed via tidal stripping of a massive progenitor because no dwarf galaxies are known to host such massive central black holes. Our result is consistent with the conclusion by \citet{Frank11} who reported that at the 96~per~cent confidence UCD3 has a black hole with less than $20$~per~cent of its stellar mass. They also found that a $5$~per~cent black hole fraction was within the 1-sigma error bars of their measurement. The 3-sigma significant black hole mass detection revealed by our analysis corresponds to about $4$~per~cent of the UCD3 total stellar mass and it is almost as massive as the central black hole in the Milky Way \citep{Gillessen09,Ghez08}. 

If we assume that the UCD3 progenitor galaxy followed the scaling relation between the black hole mass and the bulge stellar mass \citep{Kormendy13}, then its progenitor must have had a bulge of about $2\times 10^9 M_{\odot}$ which is somewhat smaller than the Milky Way bulge \citep{Haering04} and comparable to an entire mass of typical dwarf early-type galaxy. Its old age, high metallicity, supersolar [Mg/Fe] abundance ratio argue against the dwarf galaxy progenitor as those usually exhibit intermediate age populations with Solar $\alpha$/Fe abundance ratios \citep{Chilingarian09}. \citet{Pechetti+17} show that early-type progenitors in a similar mass range have central $M/L$ enhancements consistent with those seen in UCDs. Also, \citet{Nguyen+17} shows that some low luminosity early-type galaxies do host BHs; in particular Messier~32 harbours a BH with the mass similar to the one found in UCD3. 

Tidal stripping is known to act efficiently on disc galaxies: numerical simulations suggest that a Milky Way sized progenitor will lose its disc almost completely and about 90~per~cent of the total stellar mass in 300--400~Myr after the first pericentral passage at about 200~kpc from a Messier~87-like galaxy on a tangential orbit \citep{Chilingarian+09}. Closer passages to a stripping galaxy will leave behind an almost ``naked'' nucleus harbouring a massive black hole \citep{Seth14}, which can survive for long period of time without being accreted because the dynamical friction acceleration is proportional to the total mass and, therefore, is much lower for a smaller stellar system than for a larger galaxy. 

We stress that the UCD's central black hole mass alone should not be used as an estimate for the total mass of its progenitor but only of its spheroidal component. In the late-type galaxies with small bulges or no bulges at all, such as Messier~33 or Messier~101, no central black holes have been detected down to very low mass limits of 1,500~$M_{\odot}$ \citep{Gebhardt01,Kormendy07}. However, there are several rare exceptions such as NGC4395, a bulgeless galaxy hosting a low-mass central black hole about an order of magnitude less massive than what we found in UCD3 \citep{FH03, DenBrok15}.

Besides UCD3 in the Fornax cluster observed with SINFONI, there are three other massive ($10^8 M_{\odot}$) ultracompact dwarf galaxies, M60-UCD1, M59cO, and VUCD3, all in the Virgo cluster, in which central massive black holes were found using AO assisted IFU spectroscopy with NIFS at the 8-m Gemini telescope \citep{Seth14,Ahn17}. All four of them host massive black holes millions to tens of millions times more massive than the Sun. There are also two lower mass ($10^7 M_{\odot}$) UCDs in the nearby galaxy group Centaurus~A, UCD320 and UCD~330 \citep{Voggel18}, where SINFONI observations did not reveal central black holes down to a mass limit of $10^5 M_{\odot}$ (in UCD~330). 

This small sample of six UCDs across a mass range $10^7 - 10^8 M_{\odot}$ is subject to selection effects because only bright and extended UCDs allow for a sufficient spatial resolution and signal to noise to enable a central black hole detection with the currently available instrumentation. The central component in a UCD has to be brighter than $m_{V} \approx 18$~mag to be able to serve as a tip-tilt source for AO systems at 8-m class telescopes. This restricts us at the Virgo/Fornax distance to the few very bright UCDs with $M_V \approx -13$ mag. In addition, its angular extension should be of the order of an arcsecond, or otherwise we will not be able to extract enough independent measurements of radial velocities and velocity dispersions in annular/sectoral bins along the radius (AO assisted IFU spectroscopy allows one to achieve spatial resolution of about 0.1--0.2~arcsec). Therefore, only UCDs in the vicinity of the Local Group out to a distance of $\sim25$~Mpc (such as the Virgo or Fornax clusters) can be observed in sufficient detail to confidently detect a black hole. For Fornax/Virgo distance, this restricts us to UCDs with $~> 10^8 M_{\odot}$, while at the CenA distance this restricts us to masses $~> 10^7 M_{\odot}$.

All four UCDs known as of now to host SMBHs are $\alpha$-enhanced: M60-UCD1 and M59cO have [Mg/Fe]$\approx +0.2$~dex, UCD3 similarly has [Mg/Fe]$\approx +0.2$~dex and VUCD3 reaches [Mg/Fe]$\approx +0.5$~dex (derived from the analysis of Lick indices published in \citealp{Francis12}). That indicates that their stellar populations were formed during  relatively short bursts of star formation \citep{Thomas05}, which is typical for large bulges and elliptical galaxies. On the other hand, both Cen~A UCDs without black holes \citep{Voggel18} have Solar $\alpha$/Fe ratios similar to those observed in nuclei of dwarf galaxies \citep{Chilingarian09,Paudel10}. 

Within the still limited statistics, it seems that the $\alpha$-enhancement could become a secondary indicator for the black hole presence in UCDs in the framework of the tidal stripping scenario: UCDs originating from progenitor galaxies with massive bulges, which are always dominated by $\alpha$-enhanced populations, should host black holes, whereas UCDs originating from dwarf elliptical galaxies or larger late-type discs with small bulges, which normally have solar $\alpha$/Fe abundance ratios should be black hole free. This argument applies only to tidally stripped systems, which we believe the brightest UCDs are, and one has to keep in mind that metal-rich globular clusters having a different origin are also often $\alpha$-enhanced.

\citet{Paudel10} provided [$\alpha$/Fe] measurements in a sample of 10 UCDs where five of them turned out to be strongly $\alpha$-enhanced ([$\alpha$/Fe]$>+0.2$~dex). From the remaining five UCDs two are intermediate objects ($0.0<[\alpha$/Fe$]<+0.2$~dex) and three possess $\alpha$-element abundances close to the Solar value ($[\alpha/Fe]\leq 0$~dex). However, the vast majority (17/19) of luminous ($M_V < -10.5$~mag) Virgo and Fornax cluster UCDs presented in \citet{Francis12} are $\alpha$-enhanced \citep[see also][]{BRSF11,Sandoval+15}.

Another feature worth mentioning is that all 4 UCDs with detected black holes possess two component surface brightness profiles. The effective radius of the inner component in all 4 galaxies is about $10$~pc while the light fraction compared to the whole galaxy varies. These inner components have similar properties to nuclear star clusters observed in spiral and elliptical galaxies \citep{2004AJ....127..105B,2007ApJ...665.1084B} and might well represent the nuclear clusters of their progenitors, while the outer components are the leftovers of their bulges (see the discussion in \citealp{Pfeffer13} about two-component brightness profiles in UCDs).


Hence, if we assume a connection between the presence of a massive central black hole and $\alpha$-enhancement, then we can expect that up-to 80~per~cent of all luminous UCDs host massive central black holes. This fraction should be lower in groups of galaxies compared to clusters because group central galaxies are less massive and extended than cluster dominant (cD) galaxies and, therefore, would not act as efficiently as a central body that performs tidal stripping. Hence, encounters with closer pericentral distances are needed in galaxy groups compared to clusters in order to achieve a similar degree of tidal stripping for progenitors of the same mass and morphology. In addition, the relative velocities in groups on average are substantially lower than in clusters. Therefore, a close encounter between a relatively large and dense progenitor system hosting a massive black hole with a central galaxy in a group required to achieve a 99~per~cent stellar mass loss will likely result in a merger without a remnant (i.e. a UCD) surviving it. 

Nucleated dwarf galaxies or more extended discs with lower surface densities, however, can still be efficiently stripped and have their nuclei survive the interaction \citep[see a possible example in][]{Lin+16}. A larger progenitor in a group might still lose a significant fraction of its stars and survive the encounter if it does not pass too close to the group centre. This process will lead to the formation of black hole hosting compact elliptical galaxies rather than UCDs if relatively massive progenitors are stripped. This hypothesis is supported by the fact that about 70~per~cent of 195 cEs discovered in \citet{Chilingarian15} are hosted in groups with less than 20 confirmed members. Also, the relatively rich nearby NGC~5846 group hosts two cEs, one with a confirmed supermassive black hole \citep{2008ApJ...677..238D} and another one with a central bump in the velocity dispersion map suggesting the presence of a black hole \citep{2010MNRAS.405L..11C}, although no luminous UCDs have been found there despite the availability of high quality HST imaging in the group centre. 

On the other hand, the stripping of dwarf galaxies or intermediate luminosity late-type discs will lead to the formation of UCDs or UCD-like objects without massive central black holes like those recently discovered in the Cen~A group. Our Local Group also hosts a few such objects (e.g. $\omega$~Cen, Messier~54), some of which are stripped to a higher degree and look similar to ``normal'' globular clusters (Terzan~5, NGC~6388, NGC~6441, B091-D) even though their internal structure and dynamics suggests their origin via tidal stripping \citep{Zolotukhin17}.

This discussion brings us to a conclusion that luminous metal-rich $\alpha$-enhanced UCDs, which represent a majority of luminous UCDs identified in nearby galaxy clusters (with a few notable exceptions such as the massive metal-poor VUCD~7) likely originated from massive progenitor galaxies and inherited their black holes. At the same time, more metal-poor UCDs in groups and clusters with Solar $\alpha$/Fe abundance ratios were probably formed via tidal stripping of dwarf galaxies or intermediate luminosity discs with small bulges and, hence, they are not expected to host massive central black holes. Therefore, integrated stellar $\alpha$/Fe ratios could be considered as a secondary indicator for the presence of central SMBHs in UCDs and used for statistical studies, because they are much easier to obtain than spatially resolved kinematics.




\section*{Acknowledgements}
AA and IC acknowledge the support by the Russian Science Foundation grant 17-72-20119. IC and AA are also grateful to the ESO Visiting Scientist programme. IC's research is supported by the Telescope Data Center, Smithsonian Astrophysical Observatory. K.T.V.~and A.C.S.~were supported by the National Science Foundation grant AST-1350389. A.J.R. was supported by National Science Foundation grant AST-1515084 and as a Research Corporation for Science Advancement Cottrell Scholar. JS acknowledges support from National Science Foundation  grant AST-1514763 and a Packard Fellowship.  We thank R. van den Bosch for providing us with a code to compute a MGE based on the parameters of S\'ersic models and O. Sil'chenko, M. Kurtz, F. Combes, and G. Mamon for fruitful discussions about the UCDs and their origin. We thank an anonymous referee for constructive criticism of the manuscript.



\bibliographystyle{mnras}
\bibliography{UCD3_SINFONI_paper} 







\bsp	
\label{lastpage}
\end{document}